\documentclass[10pt]{article}
\usepackage[cp866]{inputenc}
\begin{document}

\centerline {{\Large\bf The quantum character of physical fields.}}
\centerline {{\Large\bf Foundations of field theories }}

\centerline {\bf L.I. Petrova}
\centerline{{\it Moscow State University, Russia, e-mail: ptr@cs.msu.su}}
\renewcommand{\abstractname}{Abstract}
\begin{abstract}

The existing field theories are based on the properties of closed
exterior forms, which are invariant ones and correspond to conservation
laws for physical fields. Hence, to understand the
foundations of field theories and their unity, one has to know how such
closed exterior forms are obtained.

In the present paper it is shown that closed exterior forms
corresponding to field theories are obtained from the equations
modelling conservation (balance) laws for material media.
It has been developed the evolutionary method that enables one
to describe the process of obtaining closed exterior forms.

The process of obtaining closed exterior forms discloses the
mechanism of evolutionary processes in material media and shows
that material media generate,  discretely, the physical structures,
from which the physical fields are formed. This justifies the quantum
character of field theories.

On the other hand, this process demonstrates the connection between field
theories and the equations for material media and points to the
fact that the foundations of field theories must be  conditioned by the
properties of material media. It is shown that the external and internal
symmetries of field theories are conditioned by the degrees of freedom
of material media. The classification parameter of physical fields
and interactions, that is, the parameter of the unified field theory,
is connected with the number of noncommutative balance conservation laws
for material media.
\end{abstract}

\bigskip
{\large\bf Introduction}

Originally, beginning from the 17th century, the physics based on the
differential equations, which describe physical processes. However, from
the 20th century, the problem of invariant (independent of the choice
of the coordinate system) description of physical phenomena arose.
As the result, the formalisms based on the tensor, group, variational
methods, on the theories of symmetries, transformations and so on
with the basic requirement of invariance were developed in physics.

This gave rise to building the field theories that enable one to
describe physical fields and their interactions. Such theories
were based on the postulates turned out practically not to be
connected with the equations of mathematical physics, which
describe physical processes. Just the absence of such a connection
produced the emergence of the problems in field theories that are
connected with investigation of general foundations of existing
field theories, their unity, and constructing the general field
theory.

In present paper it is shown that the connection of field theories with
the equations describing physical processes in material media must lie
at the basis of the general field theory.

The investigation of the foundations of field theories has been
carried out using skew-symmetric differential forms.
Skew-symmetric differential forms, which deal with differentials
and differential expressions, can describe a conjugacy of various
operators and objects. This is of principal importance for
mathematical physics and field theories since the conjugated
objects are invariants.

The properties of existing field theories are those of the closed
exterior skew-symmetric differential forms [1,2], which are conjugated
objects and correspond to conservation laws for physical fields.
The properties of closed exterior forms explicitly or implicitly
manifest themselves essentially in all formalisms of field theory, such
as the Hamilton formalism, tensor approaches, group methods, quantum
mechanics equations, the Yang-Mills theory and others.
The gauge transformations (unitary, gradient and so on), the gauge
symmetries and the identical relations of field theories are
transformations, symmetries and relations of the theory of closed
exterior forms.

Such connection between field theories and the theory of closed exterior
forms enables one to understand the properties of field theories, which
are common for all existing field theories.

However, this does not solve the basic problems of field theories.
To understand the general foundations of field theories and their
unity, one must know how the closed exterior forms connected with
field theories are obtained.

It is known that the closed exterior forms are obtained from
differential equations provided the requirements of integrability
of these equations [1].

In the present paper it is shown, firstly, that the equations
describing (balance) conservation laws for {\it material media}
serve as the differential equations from which the closed exterior
forms related to field theories and corresponding conservation
laws for {\it physical fields} are obtained. And, secondly, it is
developed the method, which is evolutionary one hence this method
allows to find not only closed exterior forms, but also to
describe the process of obtaining closed exterior forms.

The process of obtaining closed exterior forms, on one side,
demonstrates the connection between field theories and the
equations for material media, and, on other side, discloses the
mechanism of evolutionary processes in material media and shows
that material media generate physical fields. This underlines the
fact that the foundations of field theories, namely, the theories
describing physical fields, must be conditioned by the properties
of material media.

It has been possible to carry out the investigation of general foundations
of field theories due to the skew-symmetric differential forms, which, unlike to
the exterior forms, are defined on deforming manifolds and hence they possess
the evolutionary properties. The mathematical apparatus of such forms includes
some nontraditional elements such as nonidentical relations and degenerate
transformation, and this enables one to describe the evolutionary
processes, discrete transitions, quantum jumps, and generation of
various structures.

\bigskip
In the first section the general properties of field theories are
investigated with the help of closed exterior forms. In the next section
the analysis of the equations of the balance conservation laws for material media,
which describe the state of material system and the mechanism of generating
physical structures forming physical fields. In the last section the general
foundations of field theories obtained from the analysis of the equations for
material media are discussed.

\bigskip
\section {Connection of field theories with the theory of closed exterior
forms}
\subsection*{Closed exterior forms and conservation laws.}
From the closure condition of the exterior form $\theta^p$ ($p$-form):
$$
d\theta^p=0\eqno(1.1)
$$
one can see that the closed exterior form $\theta^p$ is a conserved
quantity. This means that this can correspond to a conservation law,
namely, to some conservative quantity.
If the form is closed only on pseudostructure, i.e. this form is
a closed inexact one, the closure conditions are written as
$$
d_\pi\theta^p=0\eqno(1.2)
$$
$$
d_\pi{}^*\theta^p=0\eqno(1.3)
$$
where ${}^*\theta^p$ is the dual form.

Condition (1.3), i.e. the closure condition for dual form, specifies
the pseudostructure $\pi$. \{Cohomology, sections of cotangent
bundles, the eikonal surfaces, potential surfaces,
pseudo-Riemannian and pseudo-Euclidean spaces, and others are
examples of the pseudostructures and manifolds that are made up by
pseudostructures.\}

From conditions (1.2) and (1.3) one can see the following. The dual form
(pseudostructure) and closed inexact form (conservative quantity)
made up a conjugated conservative object that can also correspond to
some conservation law. The conservation laws, to which physical fields
are subject, are just such conservation laws.

The conservative object made up by the closed inexact exterior form and
corresponding dual form is a differential-geometrical structure.
(Such differential-geometrical structures are examples of G-structures.)
The physical structures, which made up physical
fields, and corresponding conservation laws are such
differential-geometrical structures,

\subsection*{Properties of closed exterior differential forms}
{\bf Invariance. Qauge transformations.}
It is known that the closed exact form is
the differential of the form of lower degree:
$$
\theta^p=d\theta^{p-1}\eqno(1.4)
$$
Closed inexact form is also a differential, and yet not total but
interior on pseudostructure
$$
\theta^p_\pi=d_\pi\theta^{p-1}\eqno(1.5)
$$

Since the closed form is a differential (a total one if the form is exact,
or an interior one on the pseudostructure if the form is inexact), it is
obvious that the closed form turns out to be invariant under all
transformations that conserve the differential. The unitary
transformations, the tangent, the canonical, the gradient
transformations and so on are examples of such transformations.
{\it These are gauge transformations of field theories}.

With the invariance of closed forms it is connected the covariance of
relevant dual forms.

{\bf Conjugacy. Duality. Symmetries.}
The closure of exterior differential forms is the result of conjugating
the elements of exterior or dual forms. The closure property of the
exterior form means that any objects, namely, the elements of exterior
form, the components of elements, the elements of the form differential,
the exterior and dual forms, the forms of sequential degrees  and
others, turn out to be conjugated.

With the conjugacy it is connected the duality.

The example of a duality having physical sense: the closed exterior form
is a conservative quantity corresponding to conservation law, and the
closed form (as the differential) can correspond to
a certain potential force.

The conjugacy is possible if there is one or another type of symmetry.

The gauge symmetries, which are interior symmetries of field theory and
with which gauge transformations are connected, are symmetries
of closed exterior differential forms. Symmetries
of closed dual forms are exterior symmetries of the equations of field theory.

{\bf Identical relations of exterior forms.}
Since the conjugacy is a certain connection between two operators or
mathematical objects, it is evident that, to express the conjugacy
mathematically, it can be used relations.
These are identical relations.

The identical relations express the fact that each closed exterior
form is the differential of some exterior form (with the degree less
by one). In general form such an identical relation can be written as
$$
d _{\pi}\varphi=\theta _{\pi}^p\eqno(1.6)
$$

In this relation the form in the right-hand side has to be a
{\it closed} one.

Identical relations of exterior differential forms are a mathematical
expression of various types of conjugacy that leads to closed exterior
forms.

Such relations like the Poincare invariant, vector and tensor identical
relations, the Cauchi-Riemann conditions, canonical relations, the
integral relations by Stokes or Gauss-Ostrogradskii, the thermodynamic
relations, the eikonal relations, and so on are examples of identical
relations of closed exterior forms that have either the form of relation
(1.6) or its differential or integral analogs.

\subsection*{The analysis of the properties of field theories using
closed exterior forms.}
One can see that the properties of closed exterior differential
forms correspond to the properties of field theories.
Hence, the mathematical principles of the theory of closed exterior
differential forms made up the basis of field theories that is common for
all existing field theories. (It should be emphasized that the field theories
are connected with the properties of {\it inexact} closed exterior forms.)

The connection between field theories and the theory of closed
exterior forms is primary explained by the fact that the closure
conditions of exterior and dual forms correspond to conservation
laws to which physical fields are subject. It is known that the
conservation laws for physical fields are those that state an
existence of conservative physical quantities or objects. The
physical structures, which made up physical fields, and
corresponding conservation laws are differential-geometrical
structures formed by closed exterior forms and dual ones. [Below,
using the evolutionary forms it will be shown that such physical
structures arise in material media discretely.]

The properties of closed exterior and dual forms, namely, invariance,
covariance, conjugacy, and duality, lie at the basis of the group,
structural and other invariant methods of field theories.

The nondegenerate transformations of field theory are
transformations of closed exterior forms.
These are gauge transformations for spinor, scalar, vector, and tensor
fields, which are transformations of closed ($0$-form),
($1$-form), ($2$-form) and ($3$-form) respectively.

The gauge, i.e. internal, symmetries of the field theory equations
(corresponding to the gauge transformations) are those of closed exterior
forms. The external symmetries of the equations of field theory are
symmetries of closed dual forms.

The basis of field theory operators is connected with the nondegenerate
transformations of exterior differential forms.
If, in addition to the exterior differential, we introduce
the following operators: (1) $\delta$ for transformations that convert
the form of $(p+1)$ degree into the form of $p$ degree, (2) $\delta'$
for cotangent transformations, (3) $\Delta$ for the
$d\delta-\delta d$ transformation, (4) $\Delta'$ for the $d\delta'-\delta'd$
transformation, one can write down the operators in the field
theory equations in terms of these operators that act on the exterior
differential forms. The operator $\delta$ corresponds to Green's
operator, $\delta'$ does to the canonical transformation operator,
$\Delta$ does to the d'Alembert operator in 4-dimensional space, and
$\Delta'$ corresponds to the Laplace operator.

It can be shown that the equations of existing field theories are
those obtained on the basis of the properties of the exterior form
theory.
The Hamilton formalism is based on the properties of closed exterior
form of the first degree and corresponding dual form.
The closed exterior differential form $ds=-Hdt+p_j dq_j$
(the Poincare invariant) corresponds to the field equation related to
the Hamilton system.
The Schr\H{o}dinger equation in quantum mechanics is an analog
to field equation, where the conjugated coordinates are changed by
operators. It is evident that the closed exterior form of zero degree
(and dual form) correspond to quantum mechanics. Dirac's {\it bra-} and
{\it cket}- vectors constitute a closed exterior form of zero degree [3].
The properties of closed exterior form of the second degree (and dual
form) lie at the basis of the electromagnetic field equations.
The Maxwell equations may be written as [4]
$d\theta^2=0$, $d^*\theta^2=0$, where $\theta^2=
\frac{1}{2}F_{\mu\nu}dx^\mu dx^\nu$ (here $F_{\mu\nu}$ is the strength
tensor).
Closed exterior and dual forms of the third degree correspond to the
gravitational field. (However, to the physical field of given type it can
be assigned closed forms of less degree. In particular, to the Einstein
equation [5] for gravitational field it is assigned the first degree closed
form, although it was pointed out that the type of a field with
the third degree closed form corresponds to the gravitational field.)

One can recognize that the gauge transformations as well as the symmetries and
equations of field theories are connected with closed exterior forms of
given degree. This enables one to introduce a classification of physical fields and interactions according to the
degree of closed exterior form. (If denote the degree of corresponding closed exterior forms
by $k$, the case $k=0$ will correspond to strong interaction, $k=1$ will
correspond to weak interaction, $k=2$ will correspond to electromagnetic
interaction, and
$k=3$ will correspond to gravitational interaction.) This
shows that there exists a commonness between field theories describing
physical fields of different types. The degree of closed exterior forms
is a parameter that integrates fields theories into unified field theory.

\bigskip 
Thus one can see that existing invariant field theories are based on
the properties of closed exterior forms. And such a connection
also discloses the problems of existing invariant field theories.

There are no answer to the question of  how closed inexact exterior
forms, which correspond to physical structures and reflect the
properties of conservation laws and on which properties field theories
are based, are obtained.

Below we will show that the answer to these question may be obtained
from the analysis of differential equations describing the conservation
laws for material media. These are just the equations from which the
closed exterior forms whose properties correspond to field theories are
obtained.

The evolutionary method of investigating these equations applied in
the present paper enables one to understand how physical fields are formed and
what must lie at the basis of the general field theory.
{\footnotesize [The method that enables one to find the closed exterior forms
(the invariants) had been proposed by Cartan [1]. The differential equations
are imposed by the requirement of obeying the closure condition of exterior form
made up by the derivatives of these equations (it is added the requirement that
the external form differential vanishes), and next one finds the conditions that
these requirements are satisfied (the integrability conditions). This method
enables one {\it to find } the closed exterior forms (the invariants) that
can possess the equation under consideration. However, for the evolutionary
equations of mathematical physics describing physical processes it is important
not only {\it to find} closed forms, but it is also important to know
how these forms {\it are obtained}, in other words, it is important to know
how the closure conditions of exterior forms are realized evolutionary.
For this a principally new {\it evolutionary } method is necessary.]}

\section{The equations of balance conservation laws
for material system: The evolutionary processes in
material media. Origination of physical structures}
The conservation laws for material media (material media will
be considered as material systems) are the balance
conservation laws for energy, linear momentum, angular momentum, and
mass. They are described by differential equations [6-8].
{\footnotesize [Material system is a variety of elements that have
internal structure and interact to one another. Thermodynamic and gas
dynamical systems, systems of charged particles, cosmic systems, systems
of elementary particles and others are examples of material systems.
Examples of elements that constitute the material system
are electrons, protons, neutrons, atoms, fluid particles, cosmic objects and
others. The conservation laws for material systems are balance ones.
These are conservation laws that establish a balance between the variation of
physical quantity of material system and the corresponding external
action.]}

\subsection*{Nonconjugacy of the balance
conservation law equations:  Noncommutativity of the balance
conservation laws.}
The conservation laws for material systems have a peculiarity, namely,
they are noncommutative ones. To this it points out the analysis of the
equations of the balance conservation laws. (Just the noncommutativity
of the balance conservation laws is a moving force of the evolutionary
processes in material media that lead to generation of physical fields.)

It turns out that, even without a knowledge of the concrete form
of the equations for balance conservation laws, with the help of
skew-symmetric differential forms one can see their specific
features. To carry out such an investigation, in addition to
exterior skew-symmetric forms the skew-symmetric differential
forms, which possesses the evolutionary properties (and for this
reason the author named those as "evolutionary" ones), will be
used. These are skew-symmetric differential forms, which, unlike
to exterior forms, are defined on deforming manifolds. Such
skew-symmetric differential forms have a specific feature, namely,
they cannot be closed. The differential of such form does not
vanish. This differential includes the metric form differential of
deforming manifold, which is obtained due to differentiating the
basis and is nonzero. The evolutionary form commutator,
in addition to the commutator made up by the derivatives of the
coefficients of the form itself, includes (in contrast to the
commutator of the exterior form) the metric form commutator being
nonzero.

(The role of evolutionary forms in mathematical physics and field theory
is due to the fact that they, as well as exterior forms, correspond to
conservation laws. However, these conservation laws are those not for
physical fields  but for material media.)

We will analyze the equations that describe the balance conservation
laws for energy and linear momentum.

If firstly to write down these equations in the inertial reference system
and next pass to the accompanying reference system (this system is
connected with the manifold built by the trajectories of the material
system elements), in the accompanying reference system the energy
equation will be written in the form
$$
{{\partial \psi }\over {\partial \xi ^1}}\,=\,A_1 \eqno(2.1)
$$

Here $\psi$  is the functional specifying the state of material system
(the action functional, entropy or wave function can be regarded as
examples of such a functional), $\xi^1$ is the coordinate along the
trajectory, $A_1$ is the quantity that depends on specific features of
material system and on external (with respect to local domain made up
by the element and its neighborhood) energy actions onto the system.

In a similar manner, in the accompanying reference system the
equation for linear momentum appears to be reduced to the equation of
the form
$$
{{\partial \psi}\over {\partial \xi^{\nu }}}\,=\,A_{\nu },\quad \nu \,=\,2,\,...\eqno(2.2)
$$
where $\xi ^{\nu }$ are the coordinates in the direction normal to the
trajectory, $A_{\nu }$ are the quantities that depend on the specific
features of material system and on external force actions.

Eqs. (2.1) and (2.2) can be convoluted into the relation
$$
d\psi\,=\,A_{\mu }\,d\xi ^{\mu },\quad (\mu\,=\,1,\,\nu )\eqno(2.3)
$$
where $d\psi $ is the differential
expression $d\psi\,=\,(\partial \psi /\partial \xi ^{\mu })d\xi ^{\mu }$.

Relation (2.3) can be written as
$$
d\psi \,=\,\omega \eqno(2.4)
$$
here $\omega \,=\,A_{\mu }\,d\xi ^{\mu }$ is the skew-symmetric differential
form of the first degree.

Since the balance conservation laws are evolutionary ones, the relation
obtained is also an evolutionary relation.

Relation (2.4) was obtained from the equation of the balance conservation
laws for energy and linear momentum. In this relation the form $\omega $
is that of the first degree. If the equations of the balance conservation
laws for angular momentum be added to the equations for energy and linear
momentum, this form in the evolutionary relation will be a form of the
second degree. And in  combination with the equation of the balance
conservation law for mass this form will be a form of degree 3.

Thus, in general case the evolutionary relation can be written as
$$
d\psi \,=\,\omega^p \eqno(2.5)
$$
where the form degree  $p$ takes the values $p\,=\,0,1,2,3$.
The evolutionary relation for $p\,=\,0$ is similar to that in the differential
forms, and it was obtained from the interaction of time and energy of material
system.

It could be noted that the degree $p$ is connected with the number of
interacting conservation laws that is equal to $(p+1)$.

Relation obtained from the equation of the balance conservation
laws has a specific feature, namely, this relation turns out
to be nonidentical.

To justify this we shall analyze relation (2.4). This relation proves to be
nonidentical since the left-hand side of the relation is a differential,
which is a closed form, but the right-hand side of the relation
involves the differential form $\omega$, which is unclosed evolutionary
form. The metric form commutator of the manifold, on which the form $\omega $ is defined,
is nonzero since this manifold is an accompanying, deforming, manifold.
The commutator made up by the derivatives of coefficients
$A_{\mu }$ the form $\omega $ itself is also nonzero, since the coefficients
$A_{\mu }$ are of different nature, that is, some coefficients have been
obtained from the energy equation and depend on the energetic actions, whereas
the others have been obtained from the equation for linear momentum and depend
on the force actions.

In a similar manner one can prove the nonidentity of relation (2.5).

\bigskip
The nonidentity of evolutionary relation means that the balance
conservation law equations are inconsistent (nonconjugated). This
reflects the properties of the balance conservation laws
that have a governing importance for the evolutionary processes in
material media, namely, their {\it noncommutativity}.

{\footnotesize [The nonidentity of evolutionary relation points to
the fact that {\it on the initial manifold} the equations of the
balance conservation laws are nonintegrable ones: the derivatives
of these equations do not make up the differential, that is, a
closed form which can be directly integrated. This is explained by
the fact that these equations, like any equations describing
physical processes, include nonpotential terms.]}

\subsection*{Physical meaning of the equations of balance conservation
laws: Description of the state of material system.}
{\bf Nonequilibrium state of material system.}

The evolutionary relation obtained from the equations of balance
conservation laws discloses a physical meaning of these equations --
these equations describe the state of material system.

It is evident that if the balance conservation laws be commutative,
the evolutionary relation would be identical and from that it would be possible
to get the differential $d\psi $, this would indicate  that the material system
is in the equilibrium state.

However, as it has been shown, in real processes the balance conservation laws
are noncommutative. The evolutionary relation is not identical and from this
relation one cannot get the differential $d\psi $. This means that the system
state is nonequilibrium.
That is, due to
noncommutativity of the balance conservation laws the material
system state turns out to be nonequilibrium. It is evident that
the internal force producing such nonequilibrium state is
described by the evolutionary form commutator. Everything that
gives the contribution to the commutator of the form $\omega^p $
leads to emergency of internal force. (Internal force is a force
that acts inside the local domain of material system, i.e. a
domain made up by the element and its neighborhood.)

Nonidentical evolutionary relation also describes how the state of
material system varies. This turns out to be possible due to the
fact that the evolutionary nonidentical relation is a selfvarying
one. This relation includes two objects one of which appears to be
unmeasurable.  The variation of any object of the relation in some
process leads to variation of another object and, in turn, the
variation of the latter leads to variation of the former. Since
one of the objects is an unmeasurable quantity, the other cannot
be compared with the first one, and hence, the process of mutual
variation cannot stop. This process is governed by the
evolutionary form commutator, that is, by interaction between the
commutator made up by derivatives of the form itself and by metric
form commutator of deforming manifold made up by the trajectories
of material system. (This is an exchange between quantities of
different nature, between physical quantities and space-time
characteristics.)

{\footnotesize [In essence, the evolutionary equation is a correlative
relation. When changing the terms of this relation cannot become equal to
one another, but in this case they correlate to one another. The terms
of the evolutionary form commutator in nonidentical relation also correlate to
one another.]}

Selfvariation of nonidentical evolutionary relation points to the
fact that the nonequilibrium state of material system turns out
to be selfvarying. State of material system changes but holds
nonequilibrium during this process.

\bigskip
{\bf Transition of material system from nonequilibrium state
to the locally-equilibrium state. Origination of physical structure.}

The significance of the evolutionary relation selfvariation
consists in the fact that in such a process it can be realized
conditions under which the inexact, closed {\it on pseudostructure},
exterior form is obtained from the evolutionary form. This
transition is possible only as the degenerate transformation,
namely, a transformation that does not conserve the differential.
The conditions of degenerate transformation are those that
determine the direction on which interior (only along a given
direction) differential of the evolutionary form vanishes.
These are the conditions that defines the pseudostructure, i.e.
the closure conditions of dual form, and leads to realization of
the exterior form closed on pseudostructure.

As it has been already mentioned, the differential of the evolutionary
form $\omega^p$ involved into nonidentical relation (2.5) is nonzero.
That is,
$$d\omega^p\ne 0 \eqno(2.6)$$
If the conditions of degenerate transformation are realized, it will take place
the transition

$d\omega^p\ne 0 \to $ (degenerate transformation) $\to d_\pi \omega^p=0$,
$d_\pi{}^*\omega^p=0$

The relations obtained
$$d_\pi \omega^p=0,  d_\pi{}^*\omega^p=0 \eqno(2.7)$$
are the closure conditions for exterior inexact form and dual form.
This means that
it is realized the exterior form closed on pseudostructure.

In this case on the pseudostructure $\pi$ evolutionary relation (2.5) converts
into the relation
$$
d_\pi\psi=\omega_\pi^p\eqno(2.8)
$$
which proves to be an identical relation. Since the form
$\omega_\pi^p$ is a closed one, on the pseudostructure this form
turns out to be a differential. There are differentials in the
left-hand and right-hand sides of this relation. This means that
the relation obtained is an identical one.

Here it should be emphasized that under degenerate transformation the
evolutionary form remains to be unclosed and the evolutionary relation
itself remains to be nonidentical one.
(The evolutionary form differential vanishes only on pseudostructure:
the differential, which equals zero, is an interior one,
the total differential of the evolutionary form is nonzero.)

The transition from nonidentical relation (2.5) obtained from
the balance conservation laws to identical
relation (2.8) means the following. Firstly, the emergency of the
closed (on pseudostructure) inexact exterior form (relation (2.7) and
right-hand side of relation (2.8)) points to origination of physical
structure. And, secondly, the existence of the state differential
(left-hand side of relation (2.8))
points to the transition of material system from nonequilibrium state
to the locally-equilibrium state. {\it (But in this case the total
state of the material system turns out to be nonequilibrium.)}

Identical relation (2.8) points to the fact that the origination
of physical structures is connected with the transition of material system to
the locally-equilibrium state.

The origination of physical structures in material system manifests itself
as an emergency of certain observable formations, which develop
spontaneously. Such formations and their manifestations are
fluctuations, turbulent pulsations, waves, vortices, creating
massless particles, and others. (One can see that the processes
described also explain such phenomena as turbulence, radiation and
others.)

\bigskip
{\bf Conditions of degenerate transformation: degrees of
freedom of material system.}

The conditions of degenerate transformation that lead to emergency of
closed inexact exterior form are connected with any symmetries.
Since these conditions are closure conditions of dual (metric) form,
they can be caused by symmetries of coefficients of the metric form
commutator (for example, these can be symmetrical connectednesses).

Under describing material system the symmetries are conditioned by degrees of
freedom of material system.
The translational degrees of freedom, internal degrees of freedom of the
system elements, and so on can be examples of
such degrees of freedom.

The conditions of degenerate transformation (vanishing the dual
form commutator) define the pseudostructure. These conditions specify the
derivative of implicit function, which defines the direction of pseudostructure.
The speeds of various waves are examples of such derivatives:
the speed of light, the speed of sound and of electromagnetic
waves, the speed of creating particles and so on.
It can be shown that the equations for surfaces of potential (of simple
layer, double layer), integral surfaces, equations for one,
two, \dots\ eikonals, of the characteristic and of the characteristic
surfaces, the residue equations and so on serve as the equations for
pseudostructures.

To the degenerate transformation it must correspond a vanishing of some
functional expressions, such as Jacobians, determinants, the Poisson
brackets, residues and others. Vanishing of these
functional expressions is the closure condition for dual form.

And it should be emphasized once more that {\it the degenerate
transformation is realized as a transition from the accompanying
noninertial coordinate system to the locally inertial system}.
The evolutionary form is defined in the noninertial frame of reference
(deforming manifold). But the closed exterior form created
is obtained with respect to the locally-inertial frame of reference
(pseudostructure).

\subsection*{Characteristics and classification of physical structures.
Forming pseudometric and metric spaces.}
{\bf Characteristics of physical structures.}

The physical structure is a differential-geometrical structure made up by
the dual form and closed inexact form.
This is a pseudostructure (dual form) with conservative quantity (closed
inexact form). The conservative quantities describe certain charges.

Since the physical structures are generated by material media by means
of the balance conservation laws, their characteristics are connected with
the characteristics of material systems and with the characteristics of evolutionary
forms obtained from the equations of balance conservation laws.

It was already mentioned that the pseudostructure is obtained from
the condition of degenerate transformation, which is connected
with the degrees of freedom of material system.

The total differential of evolutionary form, which holds to be nonzero,
defines two another characteristics of physical structure.

The first term of the evolutionary form differential, more exactly,
its commutator, determines the value of discrete change (the quantum),
which the quantity conserved on the pseudostructure undergoes during
transition from one pseudostructure to another. The second term of the
evolutionary form commutator determines the bending of pseudostructure.
The bending specifies the characteristics that fixes the character of
the manifold deformation, which took place before physical structure
emerged. (Spin is an example of such a characteristics).

The closed exterior forms obtained correspond to the state differential
for material system. The differentials of entropy, action, potential
and others are examples of such differentials.

As it was already mentioned, in material system the created physical
structure is revealed as an observable formation. It is evident that the
characteristics of the formation (intensity, vorticity, absolute and
relative speeds of propagation of the formation), as well as those of
created physical structure, are determined by the evolutionary
form and its commutator and by the material system characteristics.

\bigskip
{\bf Classification of physical structures.}

The connection of the physical structures with the
skew-symmetric differential forms allows to introduce a classification
of these structures in dependence on parameters that specify the
skew-symmetric differential forms and enter into nonidentical and
identical relation. To determine these parameters one has to consider
the problem of integration of the nonidentical evolutionary relation.

Under degenerate transformation from the nonidentical evolutionary
relation one obtains a relation being identical on pseudostructure.
Since the right-hand side of such a relation can be expressed in terms
of differential (as well as the left-hand side), one obtains a relation
that can be integrated, and as a result he obtains a relation with the
differential forms of less by one degree.

The relation obtained after integration proves to be nonidentical
as well.

By sequential integrating the nonidentical relation of degree $p$ (in
the case of realization of corresponding degenerate transformations
and forming the identical relation), one can get a closed (on the
pseudostructure) exterior form of degree $k$, where $k$ ranges
from $p$ to $0$.

In this case one can see that after such integration the closed (on the
pseudostructure) exterior forms, which depend on two parameters, are
obtained. These parameters are the degree of evolutionary form $p$
in the evolutionary relation and the degree of created closed
forms $k$.

In addition to these parameters, another parameter appears, namely, the
dimension of space.

What is implied by the concept ``space"?

In the process of deriving the evolutionary relation two frames of reference were
used and, correspondingly, two spatial objects. The first frame of reference
is an inertial one, which is connected with the space where material
system is situated and is not directly connected with material system.
This is an inertial space, it is a metric space. (This space is also
formed by the material systems.) The second frame of reference
is a proper one, it is connected with the accompanying manifold,
which is not a metric manifold.

While generating closed forms of sequential degrees $k=p$,
$k=p-1$, \dots, $k=0$ the pseudostructures of dimensions
$(n+1-k)$: 1, \dots, $n+1$ are obtained, where $n$ is the dimension of
inertial space. As a result of transition
to the exact closed form of zero degree {\it the metric} structure of
the dimension $n+1$ is obtained.

The parameters of physical structures generated by the evolutionary
relation depend on the degree of differential forms $p$ and $k$
and on the dimension of original inertial space $n$.

With introducing the classification by numbers $p$, $k$ and $n$ one can
understand the internal connection between various physical fields.
Since physical fields are the carriers of interactions, such
classification enables one to see the connection between
interactions.

Such a classification may be presented in the form of the table given
below. This table corresponds to elementary particles.

{\footnotesize [It should be emphasized the following. Here the concept of
``interaction" is used in a twofold meaning: an interaction of the balance
conservation laws
that relates to material systems, and the physical concept of ``interaction"
that relates to physical fields and reflects the interactions of physical
structures, namely, it is connected with exact conservation laws]}.

\bigskip
\centerline{TABLE}

{\scriptsize
\noindent
\begin{tabular}{@{~}c@{~}c@{~}c@{~}c@{~}c@{~}c@{~}}
\bf interaction&$k\backslash p,n$&\bf 0&\bf 1&\bf 2&\bf 3

\\
\hline
\hline
\bf gravitation&\bf 3&&&&
    \begin{tabular}{c}
    \bf graviton\\
    $\Uparrow$\\
    electron\\
    proton\\
    neutron\\
    photon
    \end{tabular}

\\
\hline
    \begin{tabular}{l}
    \bf electro-\\
    \bf magnetic
    \end{tabular}
&\bf 2&&&
    \begin{tabular}{c}
        \bf photon2\\
    $\Uparrow$\\
    electron\\
    proton\\
    neutrino
    \end{tabular}
&\bf photon3

\\
\hline
\bf weak&\bf 1&&
    \begin{tabular}{c}
    \bf neutrino1\\
    $\Uparrow$\\
    electron\\
    quanta
    \end{tabular}
&\bf neutrino2&\bf neutrino3

\\
\hline
\bf strong&\bf 0&
    \begin{tabular}{c}
    \bf quanta0\\
    $\Uparrow$\\
    quarks?
    \end{tabular}
&
    \begin{tabular}{c}
    \bf quanta1\\
    \\

    \end{tabular}
&
\bf quanta2&\bf quanta3

    \\
\hline
\hline
    \begin{tabular}{c}
    \bf particles\\
    material\\
    nucleons?
    \end{tabular}
&
    \begin{tabular}{c}
    exact\\
    forms
    \end{tabular}
&\bf electron&\bf proton&\bf neutron&\bf deuteron?
\\
\hline
N&&1&2&3&4\\
&&time&time+&time+&time+\\
&&&1 coord.&2 coord.&3 coord.\\
\end{tabular}
}

In the Table the names of the particles created are given. Numbers placed
near particle names correspond to the space dimension. Under the names of
particles the
sources of interactions are presented. In the next to the last row we
present particles with mass (the elements of material system) formed by
interactions (the exact forms of zero degree obtained by sequential
integrating the evolutionary relations with the evolutionary forms of
degree $p$ corresponding to these particles). In the bottom row the
dimension of the {\it metric} structure created is presented.

From the Table one can see the correspondence between the degree $k$ of
the closed forms realized and the type of interactions. Thus, $k=0$
corresponds to strong interaction, $k=1$ corresponds to weak interaction,
$k=2$ corresponds to electromagnetic interaction, and $k=3$ corresponds
to gravitational interaction.

The degree $k$ of the closed forms realized and the number $p$ connected with the
number of interacting
balance conservation laws determine the type of interactions and the type
of particles created. The properties of particles are governed by the
space dimension. The last property is connected with the fact that
closed forms of equal degrees $k$, but obtained from the evolutionary
relations acting in spaces of different dimensions $n$, are distinctive
because they are defined on pseudostructures of different dimensions
(the dimension of pseudostructure $(n+1-k)$ depends on the dimension
of initial space $n$). For this reason the realized physical structures
with closed forms of degrees $k$ are distinctive in their
properties.

The parameters $p$, $k$, $n$ can range from 0 to 3. They determine some
completed cycle. The cycle involves four levels, to each of which are assign
their own values of $p$ ($p=0,1,2,3$) and space dimension $n$.

In the Table one cycle of forming physical structures is presented.
Each material system has his own completed cycle. This distinguishes one
material system from another system. One completed cycle can serve as
the beginning of another cycle (the structures formed in the preceding
cycle serve as the sources of interactions for beginning a new cycle).
This may mean that one material system (medium) proves to be imbedded
into the other material system (medium). The sequential cycles
reflect the properties of sequentially imbedded material systems.
And yet a given level has specific
properties that are inherent characteristics of the same level in
another cycles. This can be seen, for example, from comparison of
the cycle described and the cycle in which to the exact forms
there correspond conductors, semiconductors, dielectrics, and
neutral elements. The properties of elements of the third
level, namely, of neutrons in one cycle and of dielectrics in
another, are identical to the properties of so called "magnetic
monopole" [9,10].

\bigskip
{\bf Forming pseudometric and metric spaces}

The mechanism of creating the pseudostructures lies at the basis of
forming the pseudometric surfaces and their transition into metric
spaces. 

It was shown above that the evolutionary relation of degree $p$ can
generate (in the presence of degenerate transformations) closed forms
of the degrees  $p, p-1,.., 0$. While generating closed forms of
sequential degrees  $k=p, k=p-1,.., k=0$ the pseudostructures of
dimensions $(n+1-k)$: $1, ..., n+1$ are obtained. As a result of
transition to the exact closed form of zero degree the metric structure
of the dimension $n+1$ is obtained. 

Here the following should be pointed out. Physical structures are
generated by local domains of material system. These are elementary
physical structures. By combining with one another they can form
large-scale structures making up pseudomanifolds and physical
fields.

Sections of the cotangent bundles (Yang-Mills fields),
cohomologies by de Rham, singular cohomologies, pseudo-Riemannian
and pseudo-Euclidean spaces, and others are examples of
pseudostructures and spaces that are formed by pseudostructures.
Euclidean and Riemannian spaces are examples of metric manifolds
that are obtained when going to the exact forms.

What can be said about the pseudo-Riemannian manifold and Riemannian space?

The distinctive property of the Riemannian manifold is an availability of
the curvature. This means that the metric form commutator of the third
degree is nonzero. Hence, the  commutator of the
evolutionary form of third degree ($p=3$), which involves into
the proper metric form commutator, is not equal to zero. That is,
the evolutionary form that enters into the evolutionary relation is unclosed,
and the relation is nonidentical one.

When realizing pseudostructures of the dimensions  $1, 2, 3, 4$
and obtaining the closed inexact forms of the degrees $k=3, k=2,
k=1, k=0$ the pseudo-Riemannian space is formed, and the
transition to the exact form of zero degree corresponds to the
transition to the Riemannian space (see Appendix). 

It is well known that while obtaining the Einstein equations it was
assumed that there are satisfied the conditions [4,8]: 1) the Bianchi
identity is satisfied, 2) the coefficients of connectedness are symmetric,
3) the condition that the coefficients of connectedness are the Christoffel
symbols, and 4) an existence of the transformation under which the
coefficients of connectedness vanish. These conditions are the
conditions of realization of degenerate transformations for
nonidentical relations obtained from the evolutionary nonidentical
relation  with evolutionary form of the degree $p=3$ and
after going to the identical relations. In this case to the Einstein
equation the identical relations with forms of the first degree are
assigned.

\bigskip
From the description of evolutionary processes in material media 
one can see that physical fields are generated by material media.
(And thus the causality of physical processes and phenomena is
explained.)

Here it should be emphasized that the conservation laws for material
media, i.e. the balance conservation
laws for energy, linear momentum, angular momentum, and mass, which
are noncommutative ones, play a controlling role in these processes.
This is precisely the noncommutativity of the balance conservation laws
produced by external actions onto material system, which is a moving
force of evolutionary processes leading to origination of physical
structures (to which exact conservation laws are assigned).
{\footnotesize [Noncommutativity of balance conservation laws for material media
and their controlling role in evolutionary processes
accompanied by emerging  physical structures practically
have not been taken into account in the explicit form anywhere. The
mathematical apparatus of evolutionary differential forms enables one
to take into account and to describe these points.]}

\section{On the foundations of general field theory.}
The results of the analysis of the equations of conservation laws for
material media shows the connection of physical fields with material media.

This points to the fact that the fields theories that describe physical
fields must be connected with the equations that describe material
systems.

Such a connection, which is common to all field
theories, discloses the general foundations of field theories,
their quantum character, justifies the unity of field theories and can
serve as an approach to general field theory.

The theories of exterior and evolutionary skew-symmetric
differential forms, which reflect the properties of conservation
laws for physical fields and material media, allow to disclose and
justify the general principles of field theories. In this case the
properties of closed exterior forms demonstrate these principles,
and the theory of evolutionary forms justifies this. Below we
present certain of concepts that lie at the basis of field
theories. (The results obtained with using the evolutionary forms
are italicized).

1. Physical fields are formatted by physical structures that are described
by closed inexact exterior and dual forms.
{\it Physical structures are generated by material media.}
{\it Characteristics of physical structures relate to the characteristics
of material systems.}

2. {\it The conservation laws for physical fields, on which the field
theories are based, are connected with the conservation laws for
material media (with the balance conservation laws for energy, linear
momentum, angular momentum, and mass and the analog of such law for the
time).}

3. Internal and external symmetries of field theories are those of
closed exterior and  dual forms. {\it They are conditioned by
the degrees of freedom of material media.}

4. {\it The origination of physical structures, from which physical
fields are made up, proceeds {\bf discretely} under realization of the degrees
of freedom of material systems.} This explains the quantum character of
field theories.

5. The gauge transformations of field theories are transformations of
closed exterior forms.  {\it They are connected  with the degenerate
transformations of the equations of conservation laws for material
media.}

6. {\it The constants of field theory must be connected with the
characteristics of material systems.}

7. The classification parameter of physical fields
and interactions, that is, the parameter of the unified field theory,
is the degree of closed exterior forms corresponding to conservation
laws for physical fields. {\it This parameter is connected with the number of the
equations of interacting noncommutative balance conservation laws. This
connection justifies the parameter of the united field theory.}

\bigskip
The results obtained show that when building the general field
theory it is necessary to take into account the connection of
existing field theories (which are based on the conservation laws
for physical fields) with the equations of noncommutative
conservation laws for material media (the balance conservation
laws for energy, linear momentum, angular momentum and mass
and the analog of such laws for the time, which takes into account
the noncommutativity of the time and the energy of material
system).

\bigskip
\rightline{\large\bf Appendix}

\bigskip
\centerline {\large\bf Forming pseudo-Riemannian manifold and Riemannian space}

The mechanism of emerging the physical structures (that form the physical 
fields) and the material system elements elucidates the mechanism of
forming the pseudometric and metric spaces.

While deriving the evolutionary relation there were used two spatial 
objects: the accompanying manifold (connected with the material system),  
which has no metric structure for actual processes, and the inertial 
space (not connected with the material system), which is the metric 
space. 

Assume that the initial inertial space has the dimension $n=3$. The material
system in such a space is subjected to the balance conservation laws
which equations in the accompanying frame of reference turn out to be
convoluted into the evolutionary relation with $p=n=3$:
$$d\,\psi\,= \,\omega ^3 \eqno(A.1)$$
(the degree of the form $\psi $ equals 2).

The form $\omega ^3$ is defined on the accompanying manifold, and therefore
this form is an evolutionary nonintegrable form, that is, its differential
is nonzero
$$d\omega^3\ne 0\eqno(A.2)$$
And accordingly, the commutator of the form $\omega ^3$ is nonzero. 

Realization of the pseudostructure (the element of the pseudometric
space) and an emergence  of the physical structure, which the closed metric
and exterior forms correspond to, is the transition from the unclosed
evolutionary form $\omega ^3$ to the closed exterior form $\omega '^3$
(this is connected with the degenerate transform).
It must be satisfied the following relations:
$$d_{\pi }\,\omega '^3\,=\,0\eqno(A.3)$$
$$d_{\pi }\,^*\omega \,'^3\,=\,0\eqno(A.4)$$
In the present case the degree of the closed form is $k=p=3$, and the
dimension of the pseudostructure is $m=n+1-k=p+1-k=1$.

From evolutionary relation (A.1) on the pseudostructure it follows
the relation
$$d_{\pi }\,\psi \,=\,\omega '^3 \eqno(A.5)$$
which is identical one because the closed form $\omega '^3$ can be expressed
in terms of the interior differential.

From this relation it can be defined the form $d_{\pi }\psi $
that specifies a state of the system, namely, the state differential.
(In the case under consideration this is the form of degree 3).
This corresponds to the conservation law because the differential of this form
(interior on the pseudostructure) is equal to zero.

Realization of the pseudostructure (connected with an emergence of the
physical structure and a fulfillment of the conservation law) is one of 
the exhibitions of the mechanism of forming metric spaces. It worth to 
underline that the pseudostructure is realized with respect to the 
inertial frame of reference. (The degenerate transform corresponds 
to the transition from the frame of reference connected with 
the accompanying manifold to the inertial coordinate system).

With the aim to be more clear we shall make the tensor expressions correspond
to the differential forms. We can make the tensor with $p$ bottom (covariant)
subscripts correspond to the external form of degree $p$ defined on
the differentiable manifold. As it is known, the differential of the form
of degree $p$ on the differentiable manifold is the form of degree $p+1$.
We can make the tensor with $p+1$ bottom subscripts correspond to
the differential of the form of degree $p+1$. Similarly to this we make
the tensor expression $K_{\alpha ...}$ correspond to the differential
or to the commutator of the nonintegrable evolutionary form. With this
notation the commutator of the form $\omega ^3$ can be written as
$K_{\alpha \beta \gamma \chi }$, where three first subscripts correspond to
the form degree, and the fourth subscript appears while differentiating
the form (from this point and further we shall use the Greek subscripts
for the accompanying frame of reference and Latin ones for the inertial that).
The commutator of the basic metric form, which can be denoted as
$R_{\alpha \beta \gamma \chi }$,
enters into the commutator of the nonintegrable form. We can make the
3-covariant tensors $S_{jkl}$ and $T_{jkl}$ (its divergence is equal
to zero as they
corresponds to the closed forms) correspond to the closed forms
$d_{\pi }\psi$ and $\omega '^3$
(that are formed with relevance to the inertial frame of reference). And
to the pseudostructure we can assign the 1-contravariant pseudotensor $T^i$
(which corresponds to the closed metric form, i.e. the pseudostructure)
that is dual to the tensor $T_{jkl}$:
$T^i\,=\,^*T_{jkl}\,=\, {1\over 6}\varepsilon ^{ijkl}\,T_{jkl}$ (here
$\varepsilon ^{ijkl}\,=\,e_ie_je_ke_l\varepsilon _{ijkl}$, where $\varepsilon _{ijkl}$ 
is the completely skew-symmetric unit pseudotensor).
Similarly, by
$S^i\,=\,^*S_{jkl}$ denote the tensor dual to
$S_{jkl}$. Now we introduce the tensor expressions:
$$
{\bf S}_{jkl}^i\,=\,\cases{S_{jkl} \cr S^i },\quad {\bf T}_{jkl}^i\,=\,
\cases{T_{jkl}\cr T^i}$$
$\{$These tensor expressions are not tensors with covariant and contravariant
subscripts because, firstly, they combine tensors and pseudotensors, and,
secondly, in these expressions one cannot raise up and lower subscripts
because the metric is not defined as yet$\}$.

The tensor expression ${\bf S}_{jkl}^ i$ corresponds to the state differential 
and its dual form. And to physical structure  it is assigned the tensor expression 
${\bf T}_{jkl}^ i$, which is the representation 
of Bi-Structure (forms $\omega '^3$ and $^*\omega '^3$).

With taking into account relations (A.3), (A.4), the relation (A.5) can be
written in terms of the tensor expressions as
$${\bf S}_{jkl}^i\,=\,{\bf T}_{jkl}^i\eqno(A.6)$$

Relation (A.6) shows that the physical structure emerged (which is 
obtained at the expense of external actions processed by the system), 
and the state differential and the relevant dual form are in one-to-one 
correspondence.
The tensor relation ${\bf S}_{jkl}^i$  relates to the material system
(specifies its state), whereas the tensor expression ${\bf T}_{jkl}^ i$,
which corresponds to physical structure, relates to physical field.
Relation (A.6) effects the connection between physical field and material
medium.

What is the further mechanism of forming the metric space?

While emerging the physical structure a quantity that is described
by the commutator of the evolutionary form $\omega ^3$ and acts as
an internal force transforms into the potential force that acts in the
direction transverse to the pseudostructure. (If the differential of the
form $\omega ^3$ be zero, that is, the 
commutators $R_{\alpha \beta \gamma \chi }$ and
$K_{\alpha \beta \gamma \chi }$ be equal to zero, the potential force will
be equal to zero). This potential force becomes a new source
of nonequilibrium (even without the extra external actions) and can lead
to  further forming the pseudostructures.

Since relation (A.5) is identical one, it can be integrated. Because the
form $\omega '^3$ is closed, it is the interior (on the pseudostructure)
differential of the form of less by one degree
$$\omega '^3\,=\,d_{\pi }\omega ^2\eqno(A.7)$$
From relations (A.5), (A.7) it follows the relation (below, for the 
sake of convenience, we shall indicate explicitly a degree of the form $\psi $) 
$$d_{\pi }\,\psi ^2\,=\,d_{\pi }\,\omega ^2$$
which can be integrated (within the accuracy up to the less degree forms):
$$\psi ^2\,=\,\omega ^2\eqno(A.8)$$
This is an integration of the nonidentical evolutionary relation (A.1) along
a single dimensionality that has been formed.

From relation (A.7) one can see that the differential of the form $\omega ^2$
is nonzero. The form $\omega ^2$ (of degree $p-1=2$) proves to be nonintegrable
form (its commutator is nonzero) on the manifold directions remained after
integration. To the commutator of the form $\omega ^2$ it can be assigned the
tensor expression $K_{\beta \gamma \chi }^{\alpha }$ (three bottom
subscripts is the degree of the exterior form plus 1, and single top subscript
is the pseudometric dimension formed). In this case the basic commutator
can be written in the form $R_{\beta \gamma \chi }^{\alpha }$.

Here it appears some specific feature. On the one hand, the form $\omega ^2$
obtained  turns out to be nonintegrable one, and therefore, it cannot be
expressed in terms of the differential. And on the other hand, the form
$\psi ^2$ in the left-hand side of relation (A.8) must be the state
differential.
This form must become the closed form and be expressed
in terms of the differential:
$$\psi ^2\,=\,d\,\psi ^1\eqno(A.9)$$

By comparison of relations (A.8) and (A.9), we get the relation
$$d\,\psi ^1\,\cong \,\omega ^2\eqno(A.10)$$
which cannot be identity because the form $\omega ^2$ is not expressed
through the differential.

Nonidentical relation (A.10) is a relation of the type
similar to initial relation (A.1), however it is the form
of less by one degree. We can repeat the analysis similarly to that for
relation (A.1) and get the pseudostructure of the greater by one
dimension.
By sequential integrating the nonidentical relations we can obtain
the pseudometric space. The closed exterior forms of degrees
$p,\,p-1,\,...,\,0$, which are inexact, correspond to this space.
The transition to the exact form of zero degree will correspond to
the transition to the metric space.

With application of the tensor expressions these transitions
can be schematically written in the following form:

$$d\psi \cong \omega ^3, \quad d\omega ^3 \neq 0\quad (K_{\alpha \beta \gamma \chi}\neq 0,
\,\,\,R_{\alpha \beta \gamma \chi }\neq 0) \eqno(A.11)$$
\hbox to 12cm{\dotfill }

\leftline{$m\,=\,1$}
$${\bf S}_{jkl}^i\,\ =\,{\bf T}_{jkl}^i\eqno(A.12)$$
{$$+d\,\psi\,\cong\,\omega ^2,\qquad \omega ^2\neq 0:\quad(K_{\beta \gamma \chi }^{\alpha }
\neq 0,\,R_{\beta \gamma \chi }^{\alpha }\neq 0)\eqno(A.13)$$
\hbox to 12cm{\dotfill }

\leftline{$m=2$}
$${\bf S}_{kl}^{ij} \,=\,{\bf T}_{kl}^{ij}\eqno(A.14)$$
$$d\,\psi\,\cong\,\omega ^1,\qquad \omega ^1\neq 0:\quad(K_{\gamma \chi }^{\alpha \beta }
\neq 0,\,R_{\gamma \chi }^{\alpha \beta }\neq 0) \eqno(A.15)$$
\hbox to 12cm{\dotfill }

\leftline{$m=3$}
$${\bf S}_{l}^{ijk} \,=\,{\bf T}_{l}^{ijk}\eqno(A.16)$$
$$d\,\psi\,\cong\,\omega ^0,\qquad \omega ^0\neq 0:\quad(K_{\chi }^{\alpha
\beta \gamma }
\neq 0,\,R_{\chi }^{\alpha \beta \gamma }\neq 0)\eqno(A.17)$$
\hbox to 12cm{\dotfill }

\leftline{$m=4$}
$${\bf S}^{ijkl}\,=\,{\bf T}^{ijkl}\eqno(A.18)$$
$$d\,\psi\,\cong\,\int \omega ^0,\qquad \omega ^0\neq 0:\quad(K^{\alpha \beta \gamma \chi }
\neq 0,\,R_{\alpha \beta \gamma \chi }\neq 0)\eqno(A.19)$$
\hrule
$$\psi \,=\,0$$

Line (A.11) in this scheme corresponds to the nonidentical initial evolutionary 
relation (with the evolutionary forms of degree 3). Here the inequality
$d\,\omega ^3\neq 0$ is written in terms of the tensor expressions for the
commutators: ($K_{\alpha \beta \gamma \chi}\neq 0$,
$R_{\alpha \beta \gamma \chi }\neq 0$).

The dotted line corresponds to the degenerate transform and to the transition
from the nonidentical evolutionary relation to the identical relation on the
pseudostructure of dimension $m=1$ (line (A.12)), as well as to the
nonidentical relation of less by one degree (line (A.13)). Line (A.12)
involves the identical relation in the tensor expressions (see relation (A.6)),
which corresponds to identical relation (A.5) in the differential forms.

Under the degenerate transform it is once again allowed the transition from
the nonidentical relation in line (A.13) to the identical relation on the
pseudostructure of dimension $m=2$ (line (A.14)) and to a new
nonidentical relation (line (A.15)). Similar transitions can be realized
under the degenerate transforms up to the closed inexact forms of zero
degree. The solid line corresponds to the transition to the exact form.

Realization of the pseudostructures of dimensions $(1,\,...,\,4)$ and
the closed inexact forms of degrees $(3,\,...,\,0)$ (an origination of
the physical structures \hbox{${\bf S}_{jkl}^i,\,...,
\,{\bf S}^{ijkl}$}) correspond to forming
the pseudometric manifold. The transition to the exact form corresponds to
a transition to the metric space.

And what can one say concerning  the pseudo-Riemann manifold and the Riemann
space?

As it is known, when deriving the Einstein equation [11] it was supposed that
the following conditions to be satisfied: the Bianchi identity is fulfilled,
the connectedness coefficients are symmetric ones (the connectedness coefficients 
are the Christoffel symbols), and there exists a transform under which
the connectedness coefficient becomes zero. These conditions are those of
realization of the degenerate transforms for the nonidentical evolutionary
relations (A.13), (A.15), (A.17), (A.19) and transition to
the identical relation.

If the Bianchi identities be satisfied [5], then
from the tensor expression $R_{\beta \gamma \chi }^{\alpha }$ it can be 
obtained the Riemann-Christoffel tensor $G_{jkl}^i$.

To the tensor
expression $R_{\gamma \chi }^{\alpha \beta }$ it is assigned the commutator
of the first degree metric 
form $(\Gamma _{\mu \nu }^{\rho }-\Gamma _{\nu \mu }^{\rho })$,
from which under the conditions of symmetry of the connectedness coefficients
$(\Gamma _{lk}^j-\Gamma _{kl}^j)=0$ the Ricci tensor can be found.

To the tensor expression $R_{\chi }^{\alpha \beta \gamma }$ it is assigned
the connectedness $\Gamma _{\mu \nu }^{\rho }$ from which under the condition
$\Gamma _{kl}^j=\{{j\atop kl}\}$ (the connectedness coefficients are equal 
to the Christoffel 
symbols) it can be obtained the tensor expression ${\bf S}_l^{ijk}$, which
corresponds to the Einstein tensor $S_l^k=G_l^k-{1\over 2}\delta _l^kG$
(the tensors $G_l^k$ and $G$ are obtained from the Riemann-Christoffel tensor
with taking into account the symmetry of the connectedness coefficients).
To Einstein's equation it is assigned identity (A.16) that relates the tensor 
expression ${\bf S}_l^{ijk}$ with the tensor expression ${\bf T}_l^{ijk}$
that corresponds to the energy-momentum tensor. (It is well to bear in mind
that the metric tensor was not formed as yet, and therefore the operation of
transfer of bottom and top subscripts with the help of the metric tensor
proves to be inapplicable).

To the tensor expression
$R^{\alpha \beta \gamma \chi }$ it is assigned the connectedness coefficients
that under the presence of the degenerate transform vanish, and this 
corresponds to  forming the closed (inexact) metric form of zero 
degree $g_{kl}=({\bf e}_k{\bf e}_l)$.
However, at given stage this only corresponds to forming the pseudoriemann
manifold. The transition from the closed inexact form of zero degree to the
exact form of zero degree corresponds to forming the metric (the metric tensor)
and going to the Riemann space.

1. Cartan E., Les Systemes Differentials Exterieus ef Leurs Application
Geometriques. -Paris, Hermann, 1945.

2. Schutz B.~F., Geometrical Methods of Mathematical Physics.
Cambridge University Press, Cambridge, 1982.

3. Dirac P.~A.~M., The Principles of Quantum Mechanics. Clarendon Press,
Oxford, UK, 1958.

4. Wheeler J.~A., Neutrino, Gravitation and Geometry. Bologna, 1960.

5. Tonnelat M.-A., Les principles de la theorie electromagnetique
et la relativite. Masson, Paris, 1959.

6. Clark J.~F., Machesney ~M., The Dynamics of Real Gases. Butterworths,
London, 1964.

7. Dafermos C.~M. In "Nonlinear waves". Cornell University Press,
Ithaca-London, 1974.

8. Tolman R.~C., Relativity, Thermodynamics, and Cosmology. Clarendon Press,
Oxford,  UK, 1969.

9. Dirac P.~A.~M., Proc.~Roy.~Soc., {\bf A133}, 60 (1931).

10. Dirac P.~A.~M., Phys.~Rev., {\bf 74}, 817 (1948).

11. Einstein A. The Meaning of Relativity. Princeton, 1953.

\end{document}